\documentclass[runningheads]{svmult}

\usepackage{makeidx}   
\usepackage{graphicx}  
\usepackage{subeqnar}  
\usepackage{multicol}  
\usepackage{physprbb}  
\makeindex             

\begin{document}
\title*{Light Echoes in Type Ia Supernovae}
\toctitle{Light Echoes in Type Ia Supernovae}

%
%
\titlerunning{Light Echoes in Type Ia Supernovae}
%
\author{Ferdinando Patat}
\authorrunning{Ferdinando Patat}
%
%
\institute{European Southern Observatory, K. Schwarzschild Str. 2, 85748
	Garching, Germany}

\maketitle              

\begin{abstract}
Light echoes are a promising tool to probe the environment where SNe Ia 
explode and an independent source of information on the progenitor's nature.
After giving a brief introduction to the phenomenon we review
the two known cases, i.e. SNe 1991T and 1998bu. We then 
present the results we obtained from the test case of SN~1998es in NGC~632.
This object was classified as a 1991T-like event and was affected by a strong
reddening, a fact which made it a good candidate for a light echo detection.
\end{abstract}

\section{Introduction}

In the last few years the study of light echoes in Supernovae (SNe) has become
rather fashionable, since they provide a potential tool to perform a detailed
tomography of the SN environment (Crotts \cite{crotts}, and references 
therein) and, in turn, can give important insights into the progenitor's 
nature. 

The problem has been addressed by several authors (see for example
Chevalier \cite{chev}, Schaefer \cite{schae} and Sparks \cite{sparks94})
and therefore here we will give only a brief introduction to the
light echo phenomenon in the single scattering approximation.
For this purpose, let us imagine the SN immersed in a dusty medium at a 
distance $d$ from the observer. If we then consider the SN event as a 
radiation flash, whose duration $\delta t_{SN}$ is so small that 
$c\;\delta t_{SN}\ll d$, at any given time the light echo generated by the 
SN light scattered into the line of sight is confined in an ellipsoid, whose 
foci coincide with the observer and the SN itself.
 
\begin{figure}[ht]
\begin{center}
\includegraphics[width=.55\textwidth]{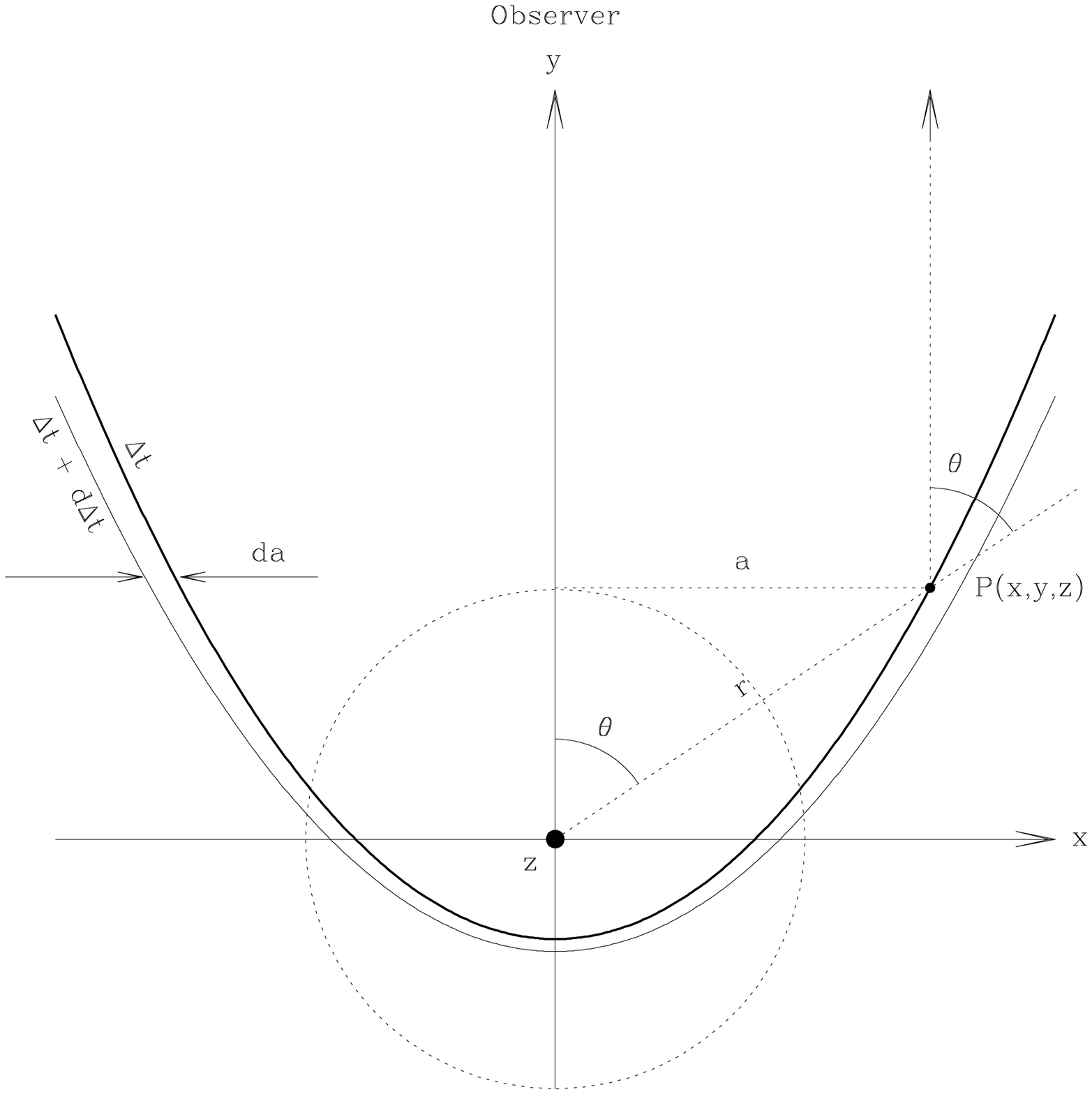}
\end{center}
\caption[]{Geometry of the problem. The $z$ axis is
perpendicular to the sheet plane.}
\label{fig:geometry}
\end{figure}

In the SN surroundings, this ellipsoid can be very well approximated
by a paraboloid, with the SN lying in its focus. If we introduce a
coordinates system centred on the source, with the $y$ axis coinciding
with the line of sight (see Fig.~\ref{fig:geometry}), then the equation of
this paraboloid can be written as

\begin{equation}
\label{eq:parabola}
y=\frac{1}{2c\Delta t} (x^2+z^2)-\frac{c\Delta t}{2}
\end{equation}

where $\Delta t$ is the time delay between the arrivals of scattered and 
direct photons, which left the SN at the same instant. It is clear that the
paraboloid can be regarded as the locus containing all scattering elements
which produce a constant $\Delta t$ and for this reason we will refer to
it as the iso--delay surface.

Let us now consider an infinitesimal volume element $dV$ lying on such
a iso-delay surface defined by a given $\Delta t$. If $L(t)$ is the number of
photons emitted per unit time by the SN, the flux of scattered photons per 
unit time and unit area which reach the observer at time $t$ with a delay 
$\Delta t$ is 

\begin{equation}
\label{eq:df}
df=\frac{L(T-\Delta t)}{4\pi r^2} \times n(r) \times a_s 
\times \omega_s \frac{\Phi(\theta)}{4\pi d^2} \times dV
\end{equation}

where $r$ is the distance of the volume element to the SN, $n(r)$ is
the number density of scattering particles, $a_s$ is the scattering
cross-section, $\omega_s$ is the dust albedo, $\theta$ is the scattering 
angle defined by $cos(\theta)=y/r$ and $\Phi(\theta)$ is the 
scattering phase function, normalized in order to have $\int_{4\pi}\Phi 
(\theta )d\Omega=1$. For the sake of simplicity, in the following we will
assume that the dust density depends only on $r=y+c\Delta t$ and that the
scattering medium is characterized by an axial symmetry around the $y$ axis.
With this simplifying hypothesis, the infinitesimal volume element can
be expressed as $dV=2\pi a\; da\;dy$, where $a$ is the impact parameter
(see Fig.~\ref{fig:geometry}). 

Then, if we set

\begin{equation}
\label{eq:f}
f(\Delta t) = \frac{c \; a_s \; \omega_s}{2}\;
\int_{-c\Delta t/2} ^{y_m} \frac{n(r) \Phi(\theta)}{r} dy 
\end{equation}

we can finally write the total scattered flux received by the observer
at time $t$ as:

\begin{equation}
\label{eq:F}
F(t)=\frac{1}{4\pi d^2}\int_0^t
L(t-\Delta t) \; f(\Delta t) \; d\Delta t
\end{equation}

while the total flux is simply $F_T(t) = L(t)/4\pi d^2 + F(t)$. From 
Eq.~\ref{eq:F} it is clear that the outcoming scattered signal is the result
of a convolution, and therefore at any given time $t$ the observer will
receive a combined signal, which contains a mixture of photons emitted
by the SN in the whole time range $0\leq\Delta t\leq t$.

As usual in this kind of studies (cfr. Chevalier \cite{chev}), we have
adopted the Henyey \& Greenstein scattering phase function
$\Phi(\theta)$, which includes the degree of forward scattering
through the parameter $g=\overline{cos(\theta)}$ ($g$=0 corresponds to an
isotropic scattering while $g$=1 indicates a complete forward scattering)
Both empirical estimates and numerical calculations (White \cite{white})
give $g\approx$0.6, while for the dust albedo we have adopted 
$\omega_s\approx$0.6 (Mathis, Rumpl \& Nordsiek \cite{mathis}).

Eq.~\ref{eq:F} can be solved numerically, using the observed light curve
of a typical Ia to get $L(t)$, but some instructive results can be obtained
analytically. In fact, the light curve of a Ia can be well approximated
by a flash ($L(t)=L_0$ for $t\leq\delta t_{SN}$ and $L(t)$=0 for 
$t>\delta t_{SN}$), and this makes the solution of Eq.~\ref{eq:F} very simple.
The flash duration can be computed from the observed light curve
as $L_0\;\delta t_{SN}=\int_0^{+\infty} L(t)\;dt$, and turns out to be of 
the order of 0.1 years for a Ia.

After normalizing to the SN luminosity at maximum, we have
$F_T(t)/F_T(0)=\delta t_{SN} \; f(t)$, so that one is left with the
solution of Eq.~\ref{eq:f} only. In general, this requires a numerical
integration, even though a few analytical solutions exist (Chevalier
\cite{chev}). The problem becomes particularly simple if one assumes
an isotropic scattering ($\Phi(\theta)=1/4\pi$) and a constant density
($n(r)=n_0$). Under these assumptions, the resulting echo luminosity has a 
very low time dependency and therefore, once the light echo radiation 
overcomes that of the SN, the light curve is expected to decline very 
slowly ($\sim$0.05 mag yr$^{-1}$) and, depending on the dust 
cloud extension, to brighten for many years. 

What is interesting to note here is that the integrated echo luminosity is 
mostly governed by the dust density and depends only mildly on the geometry
of the scattering material. For typical values of the cross section 
(6$\times$10$^{-22}$ cm$^2$) and dust density (10 cm$^{-3}$), the light echo 
is expected to take place at about 10 magnitudes below the SN maximum, and 
thus it would dominate the total luminosity starting at about two years after 
the explosion. At such epochs, the color of the object is expected to turn
blue and the spectrum to show an abrupt change from the nebular appearance
to a rather weird aspect, being a mixture of all spectra shown by the SN 
all the way back to the explosion. 
This effect, which makes the study of late phases particularly cumbersome
in such {\it unlucky} cases, was actually seen in the two known events, which 
are described in the next section.

\section{Known light echoes in Supernovae of type Ia}

The first detection of a light echo in a Ia occurred for SN~1991T (Schmidt
et al. \cite{schmidt}), which started to deviate from the usual exponential
decline at about 1.5 years past maximum, when it settled on a much slower
decline tail and showed unprecedented blue spectra. The existence of a light 
echo was definitely confirmed by the HST images taken between 1996 and 1998, 
which have indeed shown a resolved structure 
(2R$\sim$0$^{\prime\prime}$\hspace{-1.5mm}.4) evolving with time (Sparks et 
al. \cite{sparks99}). This implied the presence of dust within $\sim$50 pc 
from the SN.

The only other known case is that of SN~1998bu (Cappellaro et al. 
\cite{capp}), which so far has shown a photometric and spectroscopic
behaviour very similar to that of SN~1991T. Due to this similarity,
Cappellaro et al. \cite{capp} have modelled both light echoes using a thin
sheet of dust, perpendicular to the line of sight and located at a
distance $D$ in front of the SN. The implied optical depth was kept fixed
at the value inferred from the estimated extinction $A_V$, i.e. 0.46
for 1991T (Phillips et al. \cite{phillips}) and 0.86 for 1998bu (Jha et al. 
\cite{jha99}), so that the echo and the observed reddening are consistently
explained. In this respect, we note that the total equivalent widths $EW$
of NaI~D lines for the two objects are 1.6 \AA\/ (Patat et al. in preparation)
and 0.7 \AA\/ (Munari et al. \cite{munari}) respectively. Therefore, if
a relation between $EW$ and $A_V$ really exists, we do not have
understood it yet very well (see also Turatto, these proceedings).

With this simplifying assumptions, the solution of Eq.~\ref{eq:f} becomes
very simple and contains one single free parameter, namely the distance $D$
between the SN and the dust cloud. The reproduction of the observed data,
both spectroscopic and photometric, is fairly good (see Cappellaro et al. 
\cite{capp} and Fig.~\ref{fig:lc} here) and allows one to estimate this 
distance. In the case of SN~1991T, this turns out to be $\sim$40pc, while 
for 1998bu the inferred value is $\sim$80pc. At the distance of NGC~3368, 
this implies a diameter of about 0$^{\prime\prime}$\hspace{-1.5mm}.3, and 
therefore the light echo was expected to be resolved by HST. This was 
confirmed by WFPC2 observations performed at 762 days after maximum by 
Garnavich et al. 
\cite{garna}, who reported the presence of two main features: a ring with a 
diameter of 0$^{\prime\prime}$\hspace{-1.5mm}.48 and a barely resolved inner 
disk (2R$\sim$0$^{\prime\prime}$\hspace{-1.5mm}.14), which indicates the 
presence of scattering material within 10pc from the SN.

Now, it is quite intriguing that both SNe exploded in SAB-liner galaxies
(NGC~4527 and NGC~3368) which show and enhanced star forming rate. Moreover,
both SNe were slow decliners, even though we must say that SN~1998bu was not 
so extreme as SN~1991T and did not share its spectroscopic peculiarities. 
These facts go in the same direction of the body of evidences collected
in the last ten years concerning the possible relation between the SN
characteristics and the host galaxy morphological type (van den Bergh \&
Pazder, \cite{vdb}, Hamuy et al. \cite{hamuy96},\cite{hamuy00}, Howell
\cite{howell}). These studies have rather convincingly demonstrated that
the sub-luminous events are preferentially found in early type galaxies
(E/S0), while the super-luminous ones (1991T-like) tend to occur in
spirals (Sbc or later). What is important to underline here, is that
1991T-like events seem to be associated with young environments and are, 
therefore, the most promising candidates for the study of light echoes.
Or, in turn, if echoes are detected around such kind of SNe, this would
strengthen their association with sites of relatively recent star formation.

The sample of known light echoes is by far too small to draw any
significant conclusion and therefore we badly need to improve on the
statistics. 

\section{SN1998es in NGC~632: a test case}

The most promising case during the last years was represented by SN~1998es
in NGC~632. This SN, in fact, was classified as a 1991T-like by Jha et al.
\cite{jha}, who also noticed that the parent galaxy was classified as an S0, 
hosting a nuclear starburst (Pogge \& Eskridge
\cite{pogge}). Moreover, the SN was reported to be projected very close to
a star forming region and an high resolution spectrum showed interstellar
NaI~D features very similar to those of 1991T (Patat et al. in preparation), 
indicating the presence of a significant amount of material in front of the 
SN ($EW$=1.46 \AA).

A 1991T-like echo at the distance of NGC632 (43 Mpc, $H_0$= 75 km s$^{-1}$
Mpc$^{-1}$) would have a magnitude $B\sim$24 and therefore an 8m-class 
telescope is required. For this purpose, we have imaged NGC~632 in $B$ and $V$
with VLT-FORS2 on September 25, 2001. A careful analysis of the data did not
reveal any stellar object at the expected position. Due to the good image 
quality achieved in the 1800 sec $B$ stacked image 
(0$^{\prime\prime}$\hspace{-1.5mm}.64) and the smooth galaxy background,
we could set an upper limit for the integrated echo luminosity of $B\geq25.5$.

\begin{figure}[ht]
\begin{center}
\includegraphics[width=.65\textwidth]{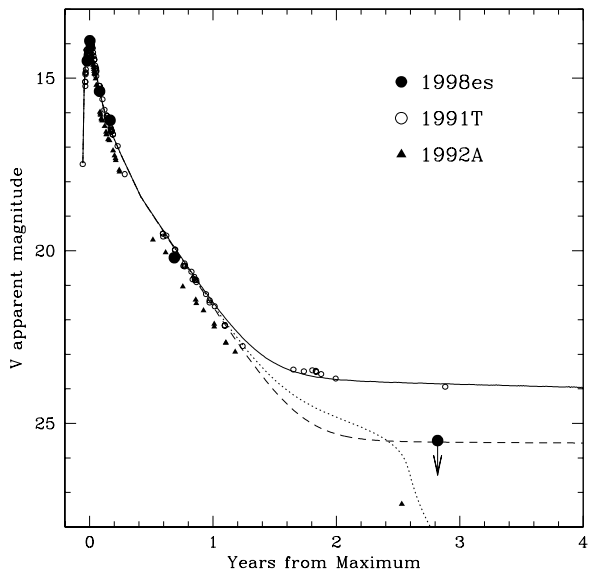}
\end{center}
\caption[]{Light curve of SN~1998es.}
\label{fig:lc}
\end{figure}

There are two possible explanations for this non-detection. The first comes
from the fact that, in the hypothesis of a thin sheet and $D\gg ct$,
the echo luminosity is proportional to $\tau/D$, where $\tau$ is 
the dust optical depth along the line of sight to the SN. Therefore, if we 
assume that $\tau$ is the same as in the case of 1991T (as the high 
resolution spectra seem to tell us), it is sufficient to place the dust sheet
$\sim$4 times further away ($D\approx$150 pc) to produce a light echo 1.5
mag fainter than in 1991T (see Fig.~\ref{fig:lc}, dashed line).
Another reasonable explanation, given the poor time coverage of SN~1998es,
is the possible confinement of the dust to a small cloud in front of the SN.
In fact, if $R$ is the radius of a spherical cloud at a distance $D$ from
the SN, the iso-delay surface will stop to intercept the scattering material
approximately for $ct\sim R^2/2D$. For example, for $D$=120 lyr and $R$=25 
lyr, the echo would last 2.6 years only. In Fig.~\ref{fig:lc} we have
plotted the numerical solution one obtains in this case (dotted line). 
We remark that, unfortunately, it is impossible to put meaningful constraints
for $D$ and $R$, due to the very poor photometric coverage at late phases.

The FORS2 images were very deep, the combined $B$ frames reaching a 
5$\sigma$ peak limiting magnitude of $\sim$27 for unresolved objects.
Despite the fact that such deep data did not show the light echo, they
led to the serendipitous discovery of a very faint tidal tail (Patat,
Carraro \& Romaniello \cite{patat}), which extends for at least 50 kpc.
This discovery strongly supports the suggestion made by Chitre \& Joshi
\cite{chitre} that NGC~632 is probably the result of a galaxy merger,
which in turn is the responsible for the observed nuclear/circum-nuclear
starburst.

Light echoes have several applications, which should be exploited
in the future. Being generated by reflections, they are strongly
polarized, with a maximum in the polarization degree at $\theta=\pi$/2.
This, in principle, can give a direct estimate of the distance 
(Sparks \cite{sparks94}). The HST light echo follow-up also allows one to
determine the distance to the SN, provided that the geometry of the
dust distribution is to some extent known.

Finally, and in our opinion most important of all, more sophisticated
models applied to high quality photometric and spectroscopic data can
provide useful information on the dust properties, its distribution around
the progenitor and possibly give some hints on its nature.

%

\end{document}